\documentclass[jeosman,preprint,fleqn,showpacs,showkeys]{revtex4}
\usepackage[pdftex]{graphicx}
\usepackage{amssymb,amsmath,bm}

\begin{document}

\title{Beyond the Kinetic Chain Process in a Stroke using a Triple Pendulum Model.}

\author{  Sun-Hyun Youn \footnote{E-mail: sunyoun@jnu.ac.kr, fax: +82-62-530-3369}}
\address{Department of Physics, Chonnam National University, Gwangju 61186, Korea}

\begin{abstract}

The efficient way to
   transfer input potential energy to the kinetic energy of a racket or bat was analyzed by
   two coupled harmonic triple pendulums. We find the most
   efficient way to transfer energy based on the kinetic chain process.
   Using control parameters, such as the release times, lengths and masses of the triple
   pendulum,
 we optimize the kinetic chain process. We also introduce a new method to get an efficient way to
 transfer initial energy to the kinetic energy of the third rod in the triple pendulum without time delay,
   which is considered an essential part of a kinetic chain process.



\end{abstract}


\maketitle

\section{Introduction}

The double pendulum model has applications  in sports such as golf
 \cite{Ref3,Ref4,Ref5,Ref6},    baseball \cite{RodCross05}, and
tennis \cite{RodCross11,youn2015New}. The  swing pattern utilized to
maximize the angular velocity of the hitting rod such as a racket,
bat, or club, has been analyzed on the assumption that the angular
velocity is the dominant factor in determining the speed of the
rebound ball \cite{RodCross11}.

   Recently, the tame-lagged torque  effect for the double pendulum system was studied.
The speed of the rebound ball can be increased not by applying
time-independent constant torques on the first rod and the second
rod, but by holding the racket for a short time without enforcing a
torque,
  and with subsequent application of torque. The speed of the rebound ball
   can be increased by $20 \%$ by choosing a proper delay time, with
   less input energy \cite{youn2015New}.

The Proper time in a stroke has its origin from "the summation of
speed principle (kinetic chain principle)" \cite{Bunn1972}, which is
one of the most important principles responsible for fast strokes.
Almost all hitting skills require that maximum speed be produced at
the end of a distal segment in a kinematic chain \cite{Throw492}.
The study of the timing patterns in the tennis forehand showed that
the tendency towards higher racket velocity was caused by
significantly different timing patterns of maximum angular pelvis
and trunk rotation \cite{Landlinger2010}. Many bio mechanical
researches on the strokes have been pursued that monitor and analyze
the stroke pattern of elite players
\cite{Example1,Example2,Example3,Example4}. These works usually
compared two groups of elite players and high performance players,
and showed why the speed of the racket of the elite group is higher.

   In this article, however, we try to find the most efficient way to
   transfer input potential energy to the kinetic energy of the racket or bat
   as a theoretical method.
   We studied two systems:  the two coupled harmonic
   oscillator,  and the triple pendulum. We find the most
   efficient way to transfer energy based on the kinetic chain process.

  The present paper is organized as follows: Section II,
  studies the two coupled harmonic oscillator. We analytically study  the
  most efficient way to convert initial potential energy stored in two
  springs  into the kinetic energy of the second body. One
  body($A$) attached by a spring to the wall is also connected to the other
  body($B$)  with another spring. We press the first body($A$)  to the wall while
  maintaining the relative distance between two bodies. Our trick is to
   fix the relative distance between the  two bodies for a certain time.
  After we release the first body $A$, the two bodies move together with
  fixed distance between them for a short time $\tau$,
   then  at $t=\tau$, we release the second spring;  the second body
  then absorbs extra kinetic energy from the second spring. Controlling
  the initial distance of the two springs and $\tau$, we can find the
  most efficient set that controls the initial potential energy to
  the kinetic energy of the second body $B$

  Section III introduces the triple pendulum. We numerically find the
  most efficient way to transfer the initial potential energy into the
  kinetic energy of the third mass. At first, relative angles of
 the three rods are fixed,  and only the angle of the first rod can
  change. After a short time $T_A$, we release the angle between
  the first and the second rods, then the triple pendulum becomes an
  effective double pendulum. After a short time $T_B$ again, we
  release the angle between the second and the third rods. Then the
  third body on the third rod moves and gets its maximum kinetic energy. We
  study the most efficient way to transfer initial potential
  energy into the kinetic energy of the third body on the third rod.
 Using the control parameters, such as release time, lengths and masses of the
 bodies,
 we study the kinetic chain process.

 Section IV introduces new method to get an efficient way to
 transfer potential energy to kinetic energy without delay time.
 Choosing the proper conditions, such as only the initial velocity
 of the two rods being zero, we can find the most efficient way to
 transfer energy. We also find that for the  same initial angular
 velocity for three rods, the initial energy is totally
 converted into the kinetic energy of the third mass on the third
 rod. This method need not be a time-lagged process,  which is an essential element
 in the kinetic chain process.   Section V summarizes the  main results and discusses
  the application of our results.

\section{Kinetic chain : Two coupled harmonic Oscillators}

\begin{figure}[htbp]
\centering
\includegraphics[width=5cm]{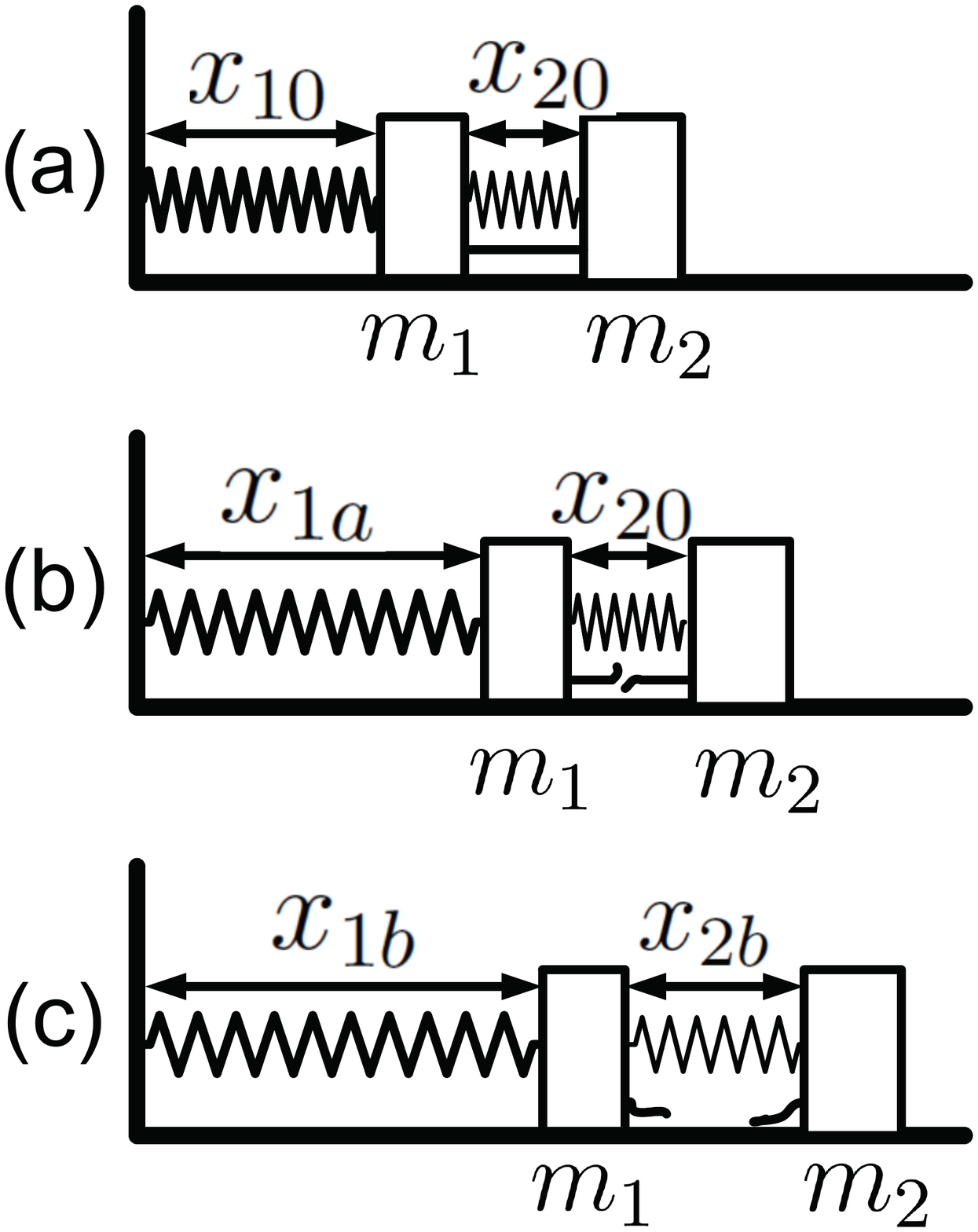}
\caption{(a) A body with mass $m_1$  starts to move from $x_1
=x_{10} $. A body with mass $m_2$ is stuck to the mass body with
mass $m_1$. (b) When the body with mass $m_1$ reaches  $x_1 =
x_{1a}$, the spring between the two bodies starts to release. (c)
Two bodies position when the body with mass $m_2$ has its maximum
velocity. } \label{FigTwoSpring}
\end{figure}

 To study the kinetic chain process, we study the two coupled harmonic
oscillator shown in Fig \ref{FigTwoSpring}. Using this system we
find  the  most efficient way to convert initial potential energy
stored in two
  springs into the kinetic energy of the second body.
  A  body($m_1$) is attached by a spring to the wall, and is also connected to an other
  body($m_2$) by another spring. We press the first body($m_1$)
   to the wall, until the distance between the first body and the
   wall   is $x_1(0) = x_{10}$, with
  keeping the relative distance between the two bodies($x_2(0)= x_{20}$).
  At this stage we tie the two bodies with a string to keep $x_{20}$ for a certain time.
  After we release the first body($m_1$), $m_1$ moves together with
  the second body($m_2$) with fixed distance between them for a short time
  $\tau$. At $t=\tau$, the position of the first body($m_1$) is $x_1 (\tau) = x_{1a} $, and
   we break the string to release the second spring; the second
   body($m_2$) then absorbs extra kinetic energy from the second spring.
If we control the initial distances $x_{1a}$ and $x_{2a}$ and
   the time $\tau$, we can find the
  most efficient set that controls the initial potential energy to
  the kinetic energy of the second body $m_2$.

  For the coupled harmonic oscillator,
we will find the general solutions. Then we impose the boundary
conditions that at time $t=0$, the second body has its maximum
kinetic energy, and no potential energy. At $t=0$, the first body is
at the equilibrium position with velocity zero, so the sum of the
kinetic energy and the potential energy for the first body at this
time is zero. In order to find $x_{1a}$, $x_{2a}$, and $\tau$, we
try to find the solution in the reverse way. Then we play back the
motion of the coupled oscillator, and find the initial conditions.

The equations for a two coupled harmonic oscillator are:
 \[   \left(
\begin{array}{cc}
m_1 & 0 \\
0 &  m_2
\end{array} \right)
\left(
\begin{array}{c}
x^{''}_{1}(t) \\
x^{''}_{2}(t)
\end{array} \right)
 = -  \left(\begin{array}{cc}
k_1  + k_2  & - k_2 \\
-k_2   &  k_2
\end{array}\right)  \left(\begin{array}{c}
x_1(t)  \\
x_2(t)
\end{array} \right),
\]
and the eigenvalue equation we have to solve is:
 \[   \left(
\begin{array}{cc}
\frac{k_1 + k_2 }{m_1} & -\frac{k_2}{m_1} \\
-\frac{k_2}{m_2} & \frac{k_2 }{m_2} \end{array} \right) \left(
\begin{array}{c}
A_1  \\
A_2
\end{array} \right)
 =  \omega^2  \left(\begin{array}{c}
A_1  \\
A_2
\end{array} \right)
\],
where, two eigenvalues $\omega^+ $ and $\omega^- $ are as follow:
\begin{eqnarray}
\omega^{\pm}  =   \sqrt {\frac{A_b \pm A_a}{2 m_1 m_2}},
\label{omegaPM}
\end{eqnarray}
The eigenvectors $\lambda^+ $ and $\lambda^- $ for $\omega^+$ and
$\omega^-$ are:
\begin{eqnarray}
\lambda^{\pm} &=& \left(
\begin{array}{c}
\lambda_1 ^{\pm}  \\
 \lambda_2 ^{\pm}
\end{array} \right) \nonumber \\
&=& \left(
\begin{array}{c}  \frac{1}{2 k_2 m_1 } (-A_c \pm A_a ) \\   1
\end{array} \right)
 , \label{lambdaPM}
\end{eqnarray}
where,
\begin{eqnarray}
A_a &=& \sqrt{(k_1 m_2 + k_2 (m_1 + m_2 ))^2 - 4 k_1 k_2 m_1 m_2
},\\
A_b  &=&  k_1 m_2 + k_2 m_2+ k_2 m_1 ,  \\
A_c  &=& k_1 m_2 + k_2 m_2 - k_2 m_1.  \label{solKabcFin}
\end{eqnarray}
Then the solutions of motion become:
\begin{eqnarray}
x_1 (t) &=& a_1 \lambda_1 ^+  \cos \omega^+ t + a_2 \lambda_1 ^+
\sin \omega^+ t +b_1 \lambda_1 ^-  \cos \omega^- t + b_2 \lambda_1
^-
\sin \omega^-   t , \nonumber \\
x_2 (t) &=& a_1 \lambda_2 ^+  \cos \omega^+ t + a_2 \lambda_2 ^+
\sin \omega^+ t +b_1 \lambda_2 ^-  \cos \omega^- t + b_2 \lambda_2
^- \sin \omega^-   t   . \label{solx1x2A}
\end{eqnarray}
If we set the boundary condition at  $t=0$,  the second body gets
its maximum velocity and the velocity of the first body is zero at
the equilibrium position. In other words, the initial potential
energy is totally converted into the kinetic energy of the second
body. For this requirement, the initial conditions should be  $x_1
(0)=0, x_2(0)=0, x_1'(0)=0,x_2'(0)=v_0 $. Then the solutions
$x_1(t), x_2 (t) $ in Eq. \ref{solx1x2A} can be written
\begin{eqnarray}
x_1 (t) &=& \sqrt{2} k_2 m_2 v_0 (\sqrt {\frac{A_b + A_a}{m_1
m_2}} \sin[\frac{(A_b - A_a ) t}{\sqrt{2} m_1 m_2 } ] \nonumber \\
& & -\sqrt {\frac{A_b - A_a}{m_1 m_2}} \sin[\frac{(A_b + A_a ) t
}{\sqrt{2} m_1 m_2  } ] ) / (\sqrt{A_a} \sqrt{\frac{A_b - A_a }{m_1
m_2}} \sqrt{\frac{A_b + A_a
}{m_1 m_2}}), \nonumber \\
x_2 (t) &=& v_0 (( A_c + A_a  )(\sqrt{\frac{A_b + A_a}{m_1
 m_2}}
 \sin[\frac{(A_b - A_a ) t}{ \sqrt{2} m_1 m_2}  ] \nonumber \\ & &+ ((A_a -A_c )
 \sqrt{\frac{A_b - A_a }{m_1 m_2}}
 \sin[\frac{(A_b + A_a ) t }{  \sqrt{2} m_1 m_2}]
  ) /(\sqrt{2} A_a  \sqrt{\frac{A_b - A_a }{m_1 m_2}}
  \sqrt{\frac{A_b + A_a}{ m_1 m_2}
  }),
   \label{solx1x2}
\end{eqnarray}
and the velocities $v_1(t), v_2(t) $ are:
\begin{eqnarray}
v_1 (t) &=&  k_2 m_2 v_0 (\cos[\frac{\sqrt{A_b - A_a }}{\sqrt{2 m_1
m_2 }}t]
- \cos[\frac{\sqrt{A_b + A_a }}{\sqrt{2 m_1 m_2 }}t] )/A_a, \nonumber \\
v_2 (t) &=& v_0 (( A_c + A_a )  \cos [ \frac{\sqrt{A_b -
A_a}}{\sqrt{2 m_1
 m_2}}t] \nonumber \\
  & &+
 (A_a- A_c   )  \cos [ \frac{\sqrt{A_b +
A_a}}{\sqrt{2 m_1
 m_2}}t])  /(\sqrt{2} A_a).    \label{solv1v2}
\end{eqnarray}

 Returning to our problem, from these solutions we
can find the release time
 of the second body. First, the second body was stuck to the
 first body with initial potential energy $ \frac{1}{2} k_2 x_{20}
 ^2 $; then, at a certain time $t= -\tau $, the second spring starts
to play a role, and at time $t=0$, the second body has its maximum
kinetic energy. To find the time $t= -\tau $, we set these two
velocities in Eq. \ref{solv1v2} to  be the same.

However, it is impossible to find the general analytic solution for
$\tau$ such that  $v_1 (-\tau) $ is equal to  $v_2 (-\tau) $.
\begin{figure}[htbp]
\centering
\includegraphics[width=5cm]{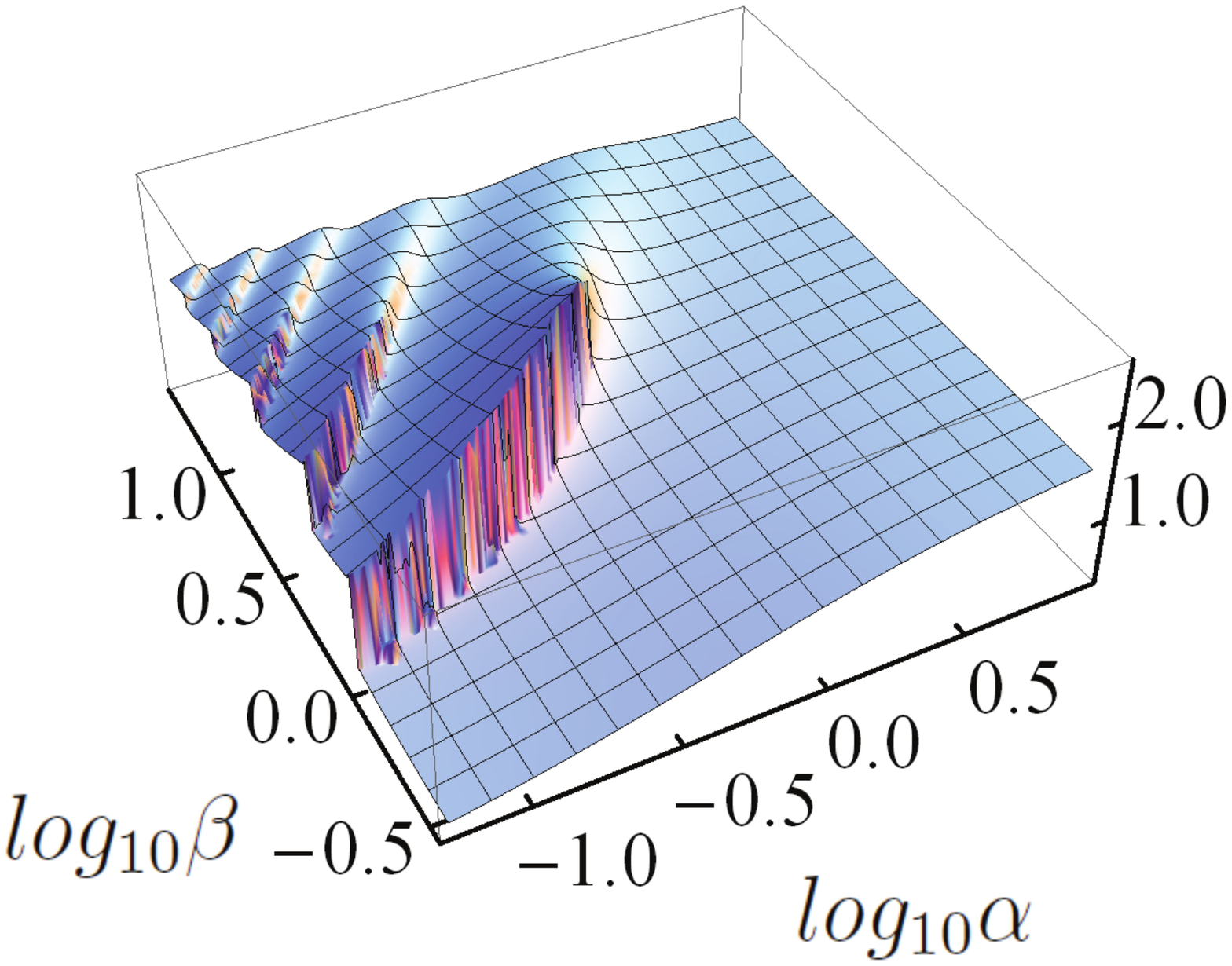}
\caption{Release time to get an maximum conversion efficiency as a
function of $log_{10} \alpha $ and $log_{10} \beta $ , where
 $\alpha = \frac{m_1}{m_2}$ and   $\beta = \frac{k_1}{k_2} $ }
\label{FReleaseTime}
\end{figure}

In Fig. \ref{FReleaseTime}, we numerically solve these equations and
plot the time $\tau$ such that $v_1 (-\tau )= v_2 (-\tau) $ as a
function of  $log_{10} \alpha $ and $log_{10} \beta $ , where
 $\alpha = \frac{m_1}{m_2}$ and   $\beta = \frac{k_1}{k_2} $.
 Although we can't find the general solution, we get some solutions
 for the extreme case ,
\begin{eqnarray}
\tau &=& \sqrt{\frac{m_1}{k_1+ k_2 }} \arccos [\frac{k_1}{k_2}] ,
\qquad   m2>>m1
 \nonumber \\
\tau &=&  \sqrt{\frac{m_2}{k_2 }}  \frac{\pi}{2}, \qquad   m1>>m2.
\label{soltau12}
\end{eqnarray}

Once, we find the solution for the time $\tau$ such that $ V_A = v_1
(-\tau) = v_2 (-\tau)$ in Eq. \ref{solv1v2}, we have to find the
analytic solutions for the two tied bodies connected to the first
spring.  In other words, two bodies sticking to each other move at
first with the same velocity by the first spring. The contraction
length of the second spring can be calculated as follows:
\begin{eqnarray}
\delta x_2  = x_2 (-\tau )- x_1 (-\tau ) \label{deltax2}
\end{eqnarray}
 The initial
starting time and the position of the first body are calculated from
the following general equations such that two bodies are attached at
the spring with boundary conditions such that  $ x_{two} (-\tau ) =
x_1 (-\tau) , v_{two}(-\tau ) = V_A $, where $x_{two} (t)$ and
$v_{tw0} (t) $ are position and the velocity of the two tied body ,
respectively. The solutions of $x_{two} $ and $v_{two}$ are as
follows:
\begin{eqnarray}
x_{two} (t) &=& x_{1}( -\tau) \cos (\sqrt{\frac{k_1}{m_1 + m_2} }
(t+\tau ) + \sqrt {\frac{m_1 + m_2 }{k_1 }} V_A  \sin
(\sqrt{\frac{k_1}{m_1 + m_2} } (t+\tau ),
\nonumber \\
v_{two} (t) &=& V_A \cos (\sqrt{\frac{k_1}{m_1 + m_2} } (t+\tau)) -
\sqrt {\frac{k_1 }{m_1 + m_2 }} x_{i} \sin (\sqrt{\frac{k_1}{m_1 +
m_2} (t+\tau } ). \label{solxvtwo}
\end{eqnarray}
 The contraction length of
the first spring can be obtained using  Eq. \ref{solxvtwo}. If we
find the time $t=-T_s $ such that $v_{two}(-T_s ) = 0$, the initial
condition of the first spring is $\delta x_1 = x_{two} (-T_s )$.

In order to transfer all the potential energy of two springs into
the kinetic energy of the second body, we tie the second spring with
contraction length $\delta x_2$ and let the two bodies move together
by the first spring with the initial position $x_{two} (-T_s)$.
After two bodies move together, at $t=- \tau$ we set the second
spring release,  then the two bodies become a two coupled harmonic
oscillator. Then at $t=0$, the kinetic energy of the second body has
its maximum value, which is the  same as the initial potential
energy of the two springs. The scenario is that the kinetic chain
process converts the initial potential energy into the kinetic
energy of the second body, by controlling the time delay.

To obtain an explicit result, we set the mass of two bodies to be
the same ($m_1 = m_2 = m =1 kg $) and the spring constants are the
same ($k_1 = k_2 =1 N/m $), then the two solutions become:
\begin{eqnarray}
x_1 (t) &=& \frac{v_0 ( \sqrt{3 +\sqrt{5}} \sin ( \sqrt{ \frac{1}{2}
(3-\sqrt{5})  }  \omega_0 t ) - \sqrt{3-\sqrt{5}} \sin ( \sqrt{
\frac{1}{2} (3+\sqrt{5}) }  \omega_0 t  )) }{\sqrt{10} \omega_0},  \nonumber \\
x_2 (t) &=& \frac{1}{2 \sqrt{10} \omega_0}\{v_0 ((1+\sqrt{5})
\sqrt{3 +\sqrt{5}} \sin ( \sqrt{ \frac{1}{2} (3-\sqrt{5})  }
\omega_0 t)
 \nonumber \\ & & - \sqrt{3-\sqrt{5}} \sqrt{\sqrt{5}-1} \sin ( \sqrt{ \frac{1}{2}
(3+\sqrt{5})  } \omega_0 t  )) \}
 \label{solx1x2FinSim}
\end{eqnarray}
\begin{eqnarray}
v_1 (t) &=& \frac{v_0 (  \cos ( \sqrt{ \frac{1}{2} (3-\sqrt{5}) }
\omega_0 t ) - \cos ( \sqrt{
\frac{1}{2} (3+\sqrt{5}) }  \omega_0 t  )) }{\sqrt{5}}, \nonumber \\
v_2 (t) &=& \frac{v_0 ((1+\sqrt{5})\cos ( \sqrt{ \frac{1}{2}
(3-\sqrt{5})  }  \omega_0 t) +\sqrt{\sqrt{5}-1} \cos ( \sqrt{
\frac{1}{2} (3+\sqrt{5})  } \omega_0 t  )) }{2 \sqrt{5}}.
 \label{solv1v2FinSim}
\end{eqnarray}
We obtain the numerical  solution for the equation $v_1 (t) = v_2(t)
$ in Eq. \ref{solv1v2FinSim}, and the result is  $ \tau \sim -1.15 s
$.
 If we want to obtain the final conditions that the second body
has a velocity and a position at $t=0$ such that $v_2(0)= 1m/s, x_2
(0) =0 $ when the first body has a velocity and a position at $t=0$
such that $v_1(0)= 0, x_1 (0) =0 $, the two bodies should  have the
same velocity at $t=-\tau$ such that $v_1(-\tau ) = v_2 (-\tau) =
0.47 m/s$. From  Eq. \ref{solx1x2FinSim}, the length of the spring
between two bodies at $t=- \tau $ should be $x_2(-\tau ) - x_1
(-\tau ) = -0.93 m -(-0.21) m = -0.72 m $.  In order to get the
velocity $v_{12}(-\tau) = 0.47 m/s$ for the two tied bodies, we can
find the initial time $T_s \sim -2 .92 s$  from  Eq. \ref{solxvtwo}.
The initial position of the first spring is $x_s = x_{two} (-T_s )
\sim -0.69 m$.

\begin{figure}[htbp]
\centering
\includegraphics[width=5cm]{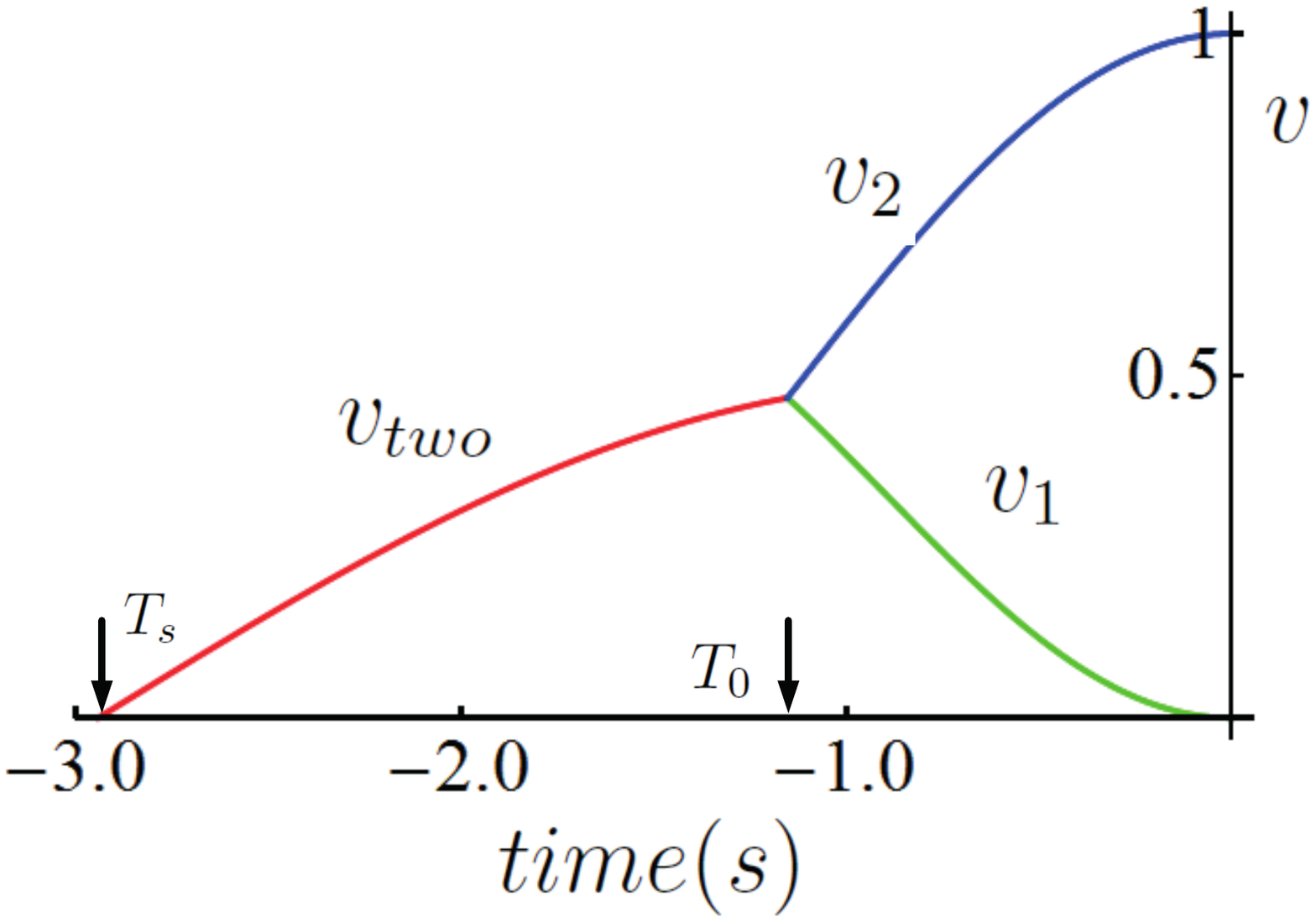}
\caption{Velocity of two bodies. Two bodies starts to move at $T_s =
-2.92 s$ at the initial position $x_s = -0.693 m$. The two bodies
moves together until  $t=T_d = -1.15 s$, and then the two bodies are
separated by the second spring. At $ t=0$, the second body has its
maximum velocity $v_2 = 1 m/s$ at $x_2(0) = 0$. The velocity of the
first body is $v_1 = 0 $ at $x_1 (0)= 0 $. We assume that $m_1 = m_2
= 1kg $, and $k_1 = k_2 = 1N/m $, for simplicity.
 }
\label{FigTwoSp}
\end{figure}

In Fig. \ref{FigTwoSp}, we plot the  velocities of the two bodies.
The two bodies start to move at $T_s = -2.92 s$ at the initial
position $x_s = -0.69 m$. The second body is tied to the first body
by the fixed spring, with keeping the relative position between two
bodies as $-0.72m $ from the equilibrium position.   The two bodies
move together until $t= -T_d = -1.15 s$, and then the two bodies are
separated by the second spring as in Fig. \ref{FigTwoSpring}. At $
t=0$, the second body has its maximum velocity $v_2 = 1 m/s$ at
$x_2(0) = 0$. The velocity of the first body is $v_1 = 0 $ at $x_1
(0)= 0 $.

\begin{figure}[htbp]
\centering
\includegraphics[width=5cm]{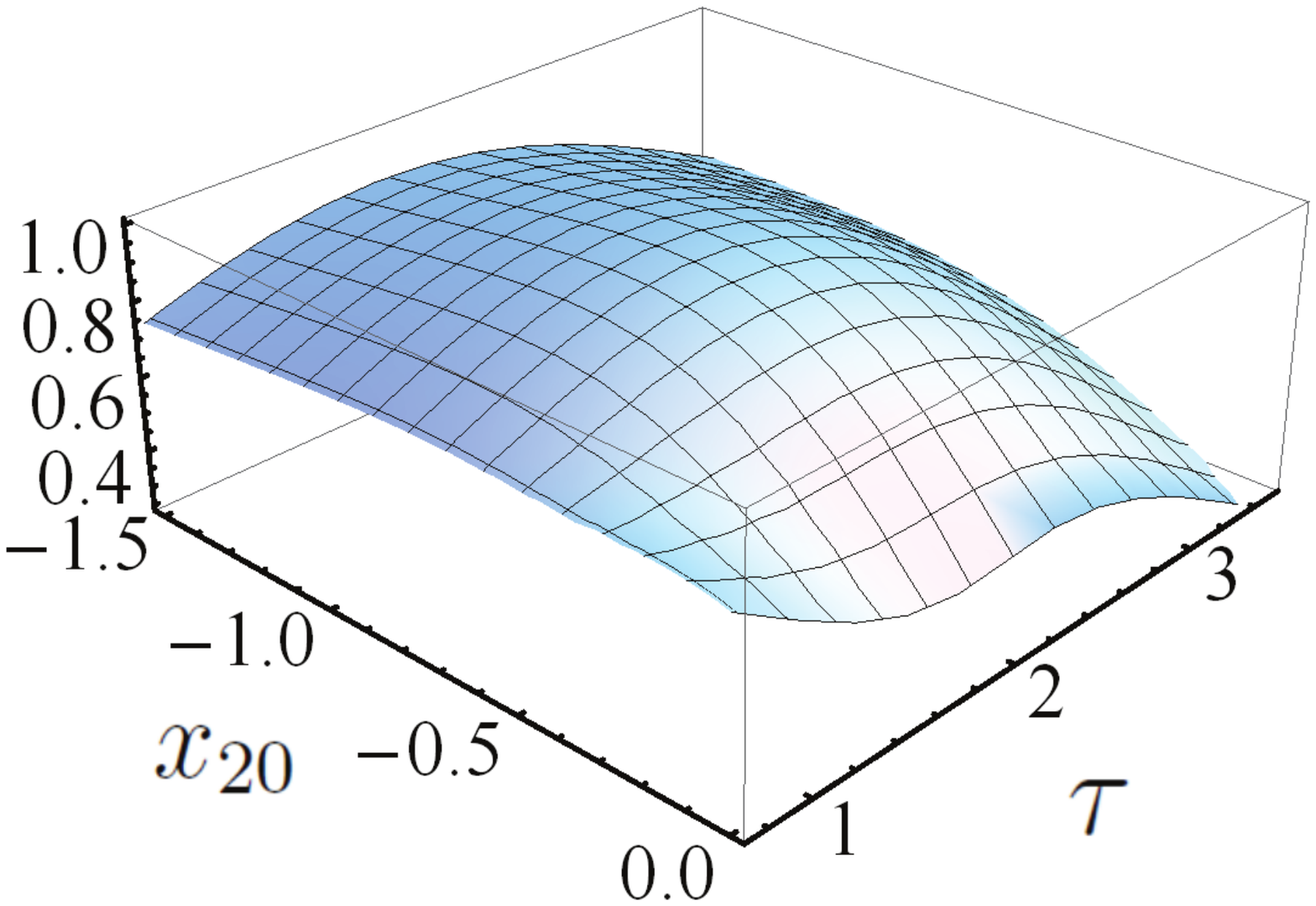}
\caption{The ratio between the maximum kinetic energy of the body
($m_2 $) and the initial potential energy stored in two springs. The
ratio is shown as a function of the release time $ \tau$ and length
($x_{20}$) of the spring between the two bodies.
 }
\label{FigEff}
\end{figure}

Up until now, we have found the solution that the initial potential
energy stored in the two springs is totally converted to the kinetic
energy of the second body at $t=0$. If we change the release time
$\tau$ and the potential energy stored in t the second spring
$x_{20}$, the kinetic energy of the second body at $t=0$ changes.
Figure \ref{FigEff} plots the ratio between the maximum kinetic
energy of the second body and the initial potential energy stored in
the two springs as a function of the release time $ \tau$ and length
($x_{20}$) of the spring between the two bodies. The figure shows
that the efficiency is sensitive to the release time. In other
words, to get  maximum efficiency, we have to very carefully control
the release time. The release time is an essential key element in
the kinetic chain process. This effect will be discussed in the next
section in detail, when we consider the kinetic chain in the three
pendulum model.

\section{The kinetic chain in the Three pendulum model}

\begin{figure}[htbp]
\centering
\includegraphics[width=5cm]{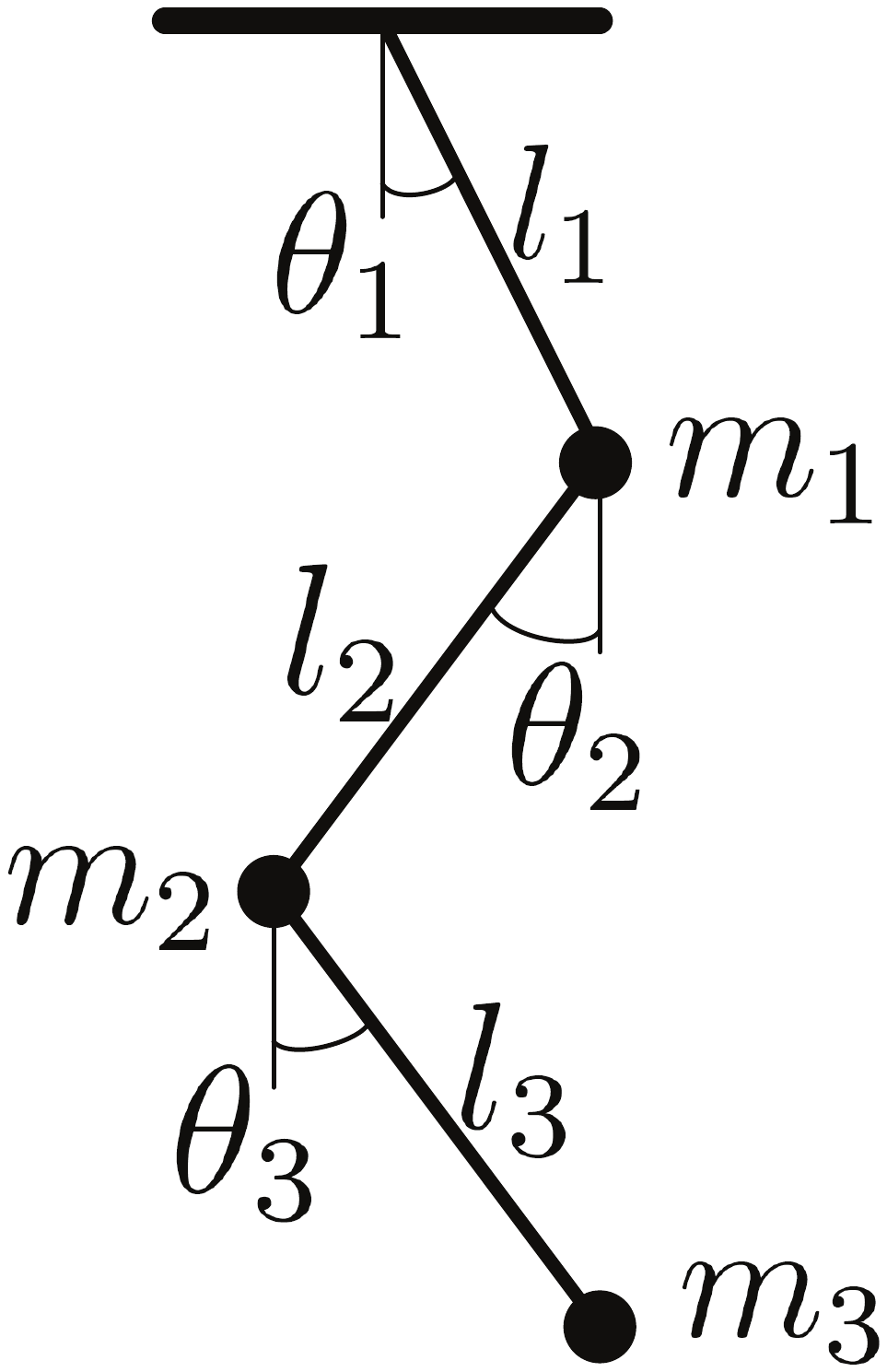}
\caption{Triple pendulum model with three massless rods and three
point masses. We assume that the forces are only the Hooke's forces
caused by the angles $\theta_1 , \theta_2, \theta_3 $ are $- k_1
\theta_2 $,
 $ -k_2 (\theta_2 -\theta_1)$ and  $ -k_3 (\theta_3 -\theta_2)$,
respectively. } \label{F3Pendulum}
\end{figure}
When we stroke a ball by racket or bat, the system is very
complicated and it is not easy to study the system in detail. But in
this article, we used the simplest model where we set the body, the
forearm and the arm, and the racket as a triple pendulum with
massless rod and mass attached at each rod's end.
  Figure
\ref{F3Pendulum} shows the  geometry of the triple pendulum model
for the swing of a racket. The triple pendulum is composed of three
rods $l_1 , l_2, l_3 $ and three masses $m_1, m_2, m_3$, as in Fig.
\ref{F3Pendulum}. We assumed that the three rods have no mass,  and
that the forces applied to the three rods are $- k_1 \theta_2 $,
 $ -k_2 (\theta_2 -\theta_1)$ and  $ -k_3 (\theta_3 -\theta_2)$,
respectively, where the three angles are defined as the angle of
each rod from the vertical line. Hooke's forces  are surely not a
sufficient model for the stroke system, but we used this simple
model to study the kinetic chain in this article.

We set the variables as follows:
\begin{displaymath}
\begin{array}{ll}
x_1 (t) = l_1 \sin \theta_1 , &y_1 (t) = l_1 \cos \theta_1,    \\
x_2 (t)= x_1(t) + l_2 \sin \theta_2 , &y_2 (t) = y_1 (t)+l_2 \cos \theta_2,  \\
x_3 (t) = x_2(t) + l_3 \sin \theta_3 , &y_3 (t) = y_3 (t)+l_3 \cos
 \theta_3, \\  \label {EqXYs}
\end{array}
\end{displaymath}
Then the kinetic energy ($T$) and the potential energy ($V$) become:
\begin{eqnarray}
T &=&\frac{1}{2} [l_1^2 (m_1+m_2+4 m_3) \beta_3'(t)^2
    +l_2^2 (m_2+m_3)  (\beta _1'(t)+\beta _3'(t))^2
     \nonumber \\ &+& l_3^2 m_3 (\beta_1'(t)+\beta _2'(t) +\beta_3'(t))^2
    +4 l_1 l_3 m_3 \cos (\beta _1(t)+\beta_2(t))
    (\beta _1'(t)+\beta_2'(t)+\beta _3'(t)) \beta_3'(t)
    \nonumber \\ &+& 2 l_2 (\beta _1'(t)  +\beta_3'(t))
    \{ l_1 (m_2+2 m_3) \cos\beta _1(t) \beta _3'(t)
    \nonumber\\  &+& l_3 m_3 \cos \beta _2(t)
     (\beta _1'(t)+\beta_2'(t) +\beta _3'(t))\}
     ],   \label {EqT}
\end{eqnarray}
\begin{eqnarray}
V = \frac{1}{2} \{ k_1 \beta_1 (t) ^2 +k_2 \beta_2 (t) ^2 + k_3
\beta_3 (t) ^2 \},
  \label {EqV}
\end{eqnarray}
 where,
\begin{eqnarray}
\beta_1 &=& \theta_2 - \theta_1, \nonumber \\
\beta_2 &=& \theta_3 - \theta_2, \nonumber \\
\beta_3 &=&  \theta_1.   \label {EqBeta}
\end{eqnarray}
Then using the Lagrangian   $ L =  T-V  $, we can find the Lagrange
Equations for $\beta_1(t), ~\beta_2 (t), ~\beta_3 (t) $, as follows:
\begin{eqnarray}
\frac{d}{dt}\frac{dL}{d \beta_i' (t)} &=& \frac{dL}{d \beta_i (t)},
i=1,2,3
 \label {EqBetaLag}
\end{eqnarray}

In order to find the efficient method to convert the initial
potential energy into the kinetic energy of $m_3$ using the kinetic
chain process, we follow the same way as in the two coupled springs
in section II based on the kinetic chain process. First, we set the
initial relative angle of the three rods as $\theta_{10},
\theta_{20}, \theta_{30}$, then we fix the initial potential energy
of this system, and then we set the three pendulum moves. But, at
the first stage, we do not release the second ($\theta_2$)  and the
third angle ($\theta_3$). With fixing two angles we solve the
effective single pendulum. After a certain time $T_A$ passes, we
release the second angle $\theta_2$, while still keeping the third
angle $\theta_3$ fixed.  After passing another time $T_B$, we
finally release the third angle $\theta_3$, and try get the maximum
speed of the mass $m_3$.
 We are interested in the time delays $T_A$ and $T_B$,which are related
  to the kinetic chain process. However, there are no analytic
  solutions to get the most efficient method to maximize the angular speed
  of  the third rods under the given initial potential energy.

  As in Section II, we  start to find the solution by the
  time reversal method. At time $t=0$, we set the boundary conditions
  that the potential energy is totally converted into the kinetic
  energy of $m_3$.  We try to numerically
solve the differential equations in Eq. \ref{EqBetaLag} with the
boundary conditions of:
\begin{eqnarray}
\beta_1 (0) &=& \beta_2 (0)= \beta_3 (0)=0, \nonumber \\
\beta_1 '(0) &=& \beta_3 '(0)=0, \nonumber\\
\beta_2'(0) &=& \dot{\theta}_{3}. \label {EqIniConditions}
\end{eqnarray}
In our simulation, the third rod with mass $m_3$ reaches its maximum
angular velocity $\dot{\theta}_{30}$ at $t=0$ when the third rod
arrives at $\theta_3 (0)=0$. The other two rods arrive at the
equilibrium positions $\theta_1 (0)= \theta_2 (0) = 0$ with angular
velocity zero.
 At this stage, we study the time behavior of the three rods retrospectively and
 we have to find the time ($T_A$) when the angular velocities of the second rod and the third rod are
 the same, as follows:
\begin{eqnarray}
\theta_{2}'(-T_A ) = \theta_{3}'(-T_A). \label {Eq23same}
\end{eqnarray}
If we find the time $T_A$, we set new equations for $\beta_1 (t),~
\beta_3 (t)$ from  Eq. \ref{EqBetaLag} with replacement $\beta_2
'(t) = 0$ and $\beta_2 (t) = \beta_2 (-T_A) $. Then the solution is
for the double  pendulum, the first pendulum is for the first rod
and the second pendulum is for the two rods whose relative angle is
fixed. When $t<-T_A$, the second rod and the third rod move together
with the angle between them  fixed as $\beta_2 (-T_A)$. At $t=-T_A$,
the third rod starts to release from the second rod and it starts to
move and reaches its maximum velocity at $t=0$.

Before the time that the third rod releases at $t=-T_A$, the second
and the third rod move together. Now we have to find the release
time of these two rods from the first rod. The time can be obtained
from the following conditions
\begin{eqnarray}
\theta_{2}'(-T_B ) = \theta_{1}'(-T_B). \label {Eq12same}
\end{eqnarray}
If we find the time $T_B$, we set new equations for $ \beta_3 (t)$
from  Eq. \ref{EqBetaLag} with replacement $\beta_2 '(t) = 0,
\beta_1 '(t) = 0 $ and $\beta_2 (t) = \beta_2 (-T_A), \beta_1 (t) =
\beta_1 (-T_B)  $. Then the solution is for the effective single
pendulum, and the relative angles among the three rods are fixed.
When $t<-T_B$, the three rods move together with the angles among
them being  $\beta_2 (-T_A), \beta_1 (-T_B) $. At $t=-T_B$, the tied
second rod and the third rod start to release from the first rod. We
can also find the start time of the effective single pendulum by
solving these equations,
\begin{eqnarray}
\theta_{3}'(-T_S ) = 0 . \label {Eq1start}
\end{eqnarray}

If we follow the time series sequentially  from $t=-T_s$, we can see
clearly the kinetic chain process. At first the angles among the
three rods are fixed, the initial angle of the first rod is
$\theta_{10} = \beta_3 (-T_S )$ and three rods starts to move
together until $t=-T_B$. At $t=-T_B$, the second rod release from
the first rod with keeping the angle between the second and the
third rod until $t=-T_A$. At $t=-T_A$, the third rod releases from
the second rod and the third rod starts to release from the second
rod, starts to move, and reaches its maximum velocity at $t=0$.

To get an explicit result, we assume the mass and length of the
pendulum. We assume that the masses of the three bodies are $m_1 =
10 kg $, $m_2 = 3 kg $, $m_3 = 0.2 kg$ and the lengths of the three
rods are $l_1 =0.3 m $, $l_2 = 0.5 m$, and $l_3 = 0.75m $. These
values come from one example of the body, forearm and arm, and
racket system. We also assume that the spring constants of the three
rods are $k_1 = 5N/m $, $k_2 = 1N/m $, and $k_3 = 0.25 N/m$.
Although these numerical values are not actual data, we can study
the role of the kinetic chain based on these particular examples.
\begin{figure}[htbp]
\centering
\includegraphics[width=7cm]{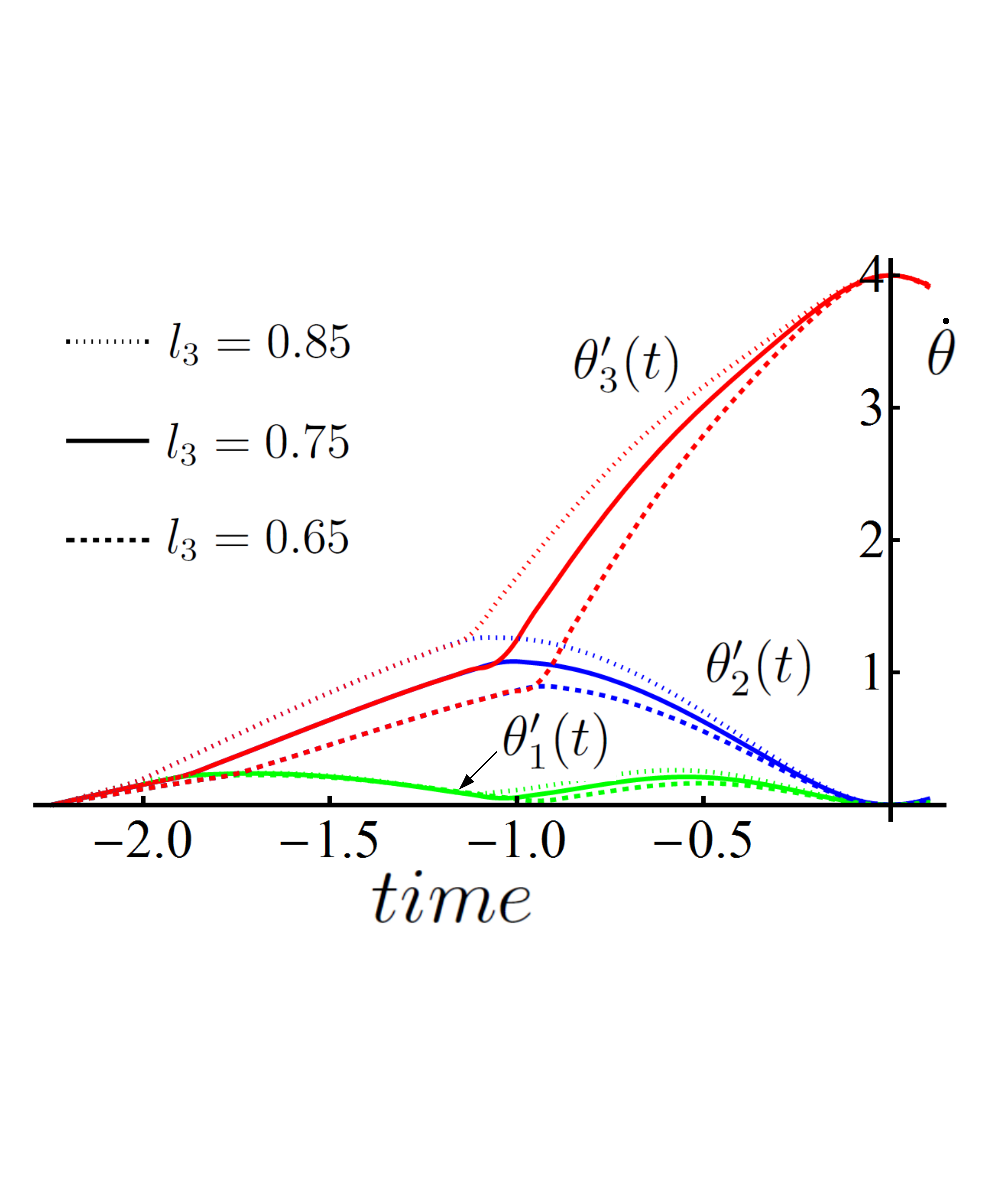}
\caption{Angular velocities of three rods. At first two rods are
moves with the same angular velocity. After a short time ($T_1$),
the second rod starts to move together with the third rods. Finally
at time $T_2$, the third rod independently starts to move.  At
$t=0$, the third rods gets maximum angular velocity as the other two
rods have zero angular velocities at equilibrium. }
\label{FRAngVelLength}
\end{figure}
Figure \ref{FRAngVelLength} shows the angular velocities of three
rods with the particular example. First, the three rods move
together with the initial angle $\theta_1=  18.3 ^{\circ} $ at $t=-
2.24 s$, until $t= -1.89 s$. At $t=-1.89 s$, the second rod is
released from the first rod. The initial relative angle between the
first and the second rod is $ 52.7 ^{\circ} $. The angular velocity
of the first rod is decreased after releasing the second rod. The
angular velocity of the second rod is increased till  $t=-1.04 s$.
Until  then, the third rod is attached to the second rod, and the
two rods move together. After $t=-1.04 s$, the three rods move
separately, and the initial relative angle between the second and
the third rod is $ 132.9 ^{\circ} $. The angular velocity of the
first rod is increased a little, and the angular velocity of the
second rod is decreased. Only the angular velocity of the third rod
is increased,  and reaches the maximum velocity at $t=0$. All the
potential energy is converted into the kinetic energy of the third
rod, which is the truly perfect kinetic chain process.

\begin{figure}[htbp]
\centering
\includegraphics[width=7cm]{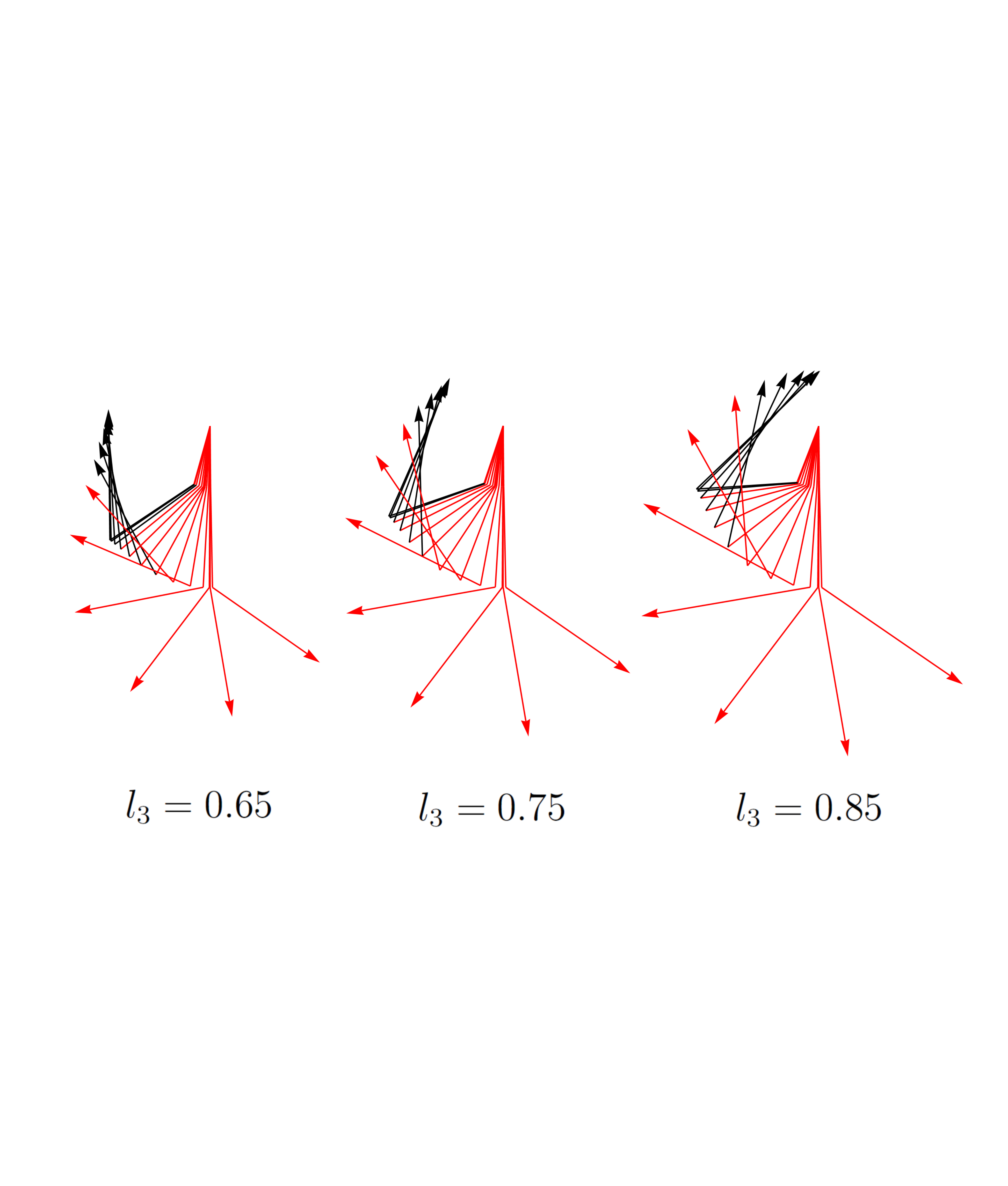}
\caption{The traces of three pendulums for different length $l_3$
with keeping other conditions unchanged. The shots are taken at
equal time intervals from the starting time. As the length $l_3$
increases the initial angle of the third rod should increase. The
red lines are under the force, and the black lines remain fixed
without generating Hooke's force. } \label{FRTraceLength}
\end{figure}

Figure \ref{FRTraceLength} shows the traces of the three pendulum
from the time $t=-2.24 s$, till $t=0.25 s$ with equal time interval
for $l_3 = 0.65 m ,~ 0.75 m,~ 0.85 m $.   The red line represents
the released rod and the black line represent the fixed rods. For
the pendulum with $l_3 =0.75 m$, at first the second and the third
rods are fixed, and the pendulum moves using the potential energy
stored at the first rod, until $t= -1.89 s$ (two steps in our
figure). After $t=-1.89 s$, the second rod is released and moves
together with the third rod till $t=-1.04 s$ (four steps in our
figure). After that, the three rods move separately, and the third
rod gets its maximum angular velocity at $t=0$.
 The start time for different $l_3$ is almost the same,  but the initial
 starting angle $\theta_1$ is increased as $l_3$ increases in Fig.
 \ref{FRTraceLength}. The initial angle is also increased as $l_3$
 changes from $l_3 =0.65 m$ to $l_3 =0.85 m$. The maximum angular
 velocity of the third rod at $t=0$ is initially set to $\dot{\theta_3} = 4
 rad/s$, but the linear velocity is increased as $l_3$ increases.
 All the potential energy is converted into the kinetic energy of the third
 rod and the kinetic energy is also increased for large $l_3$,
 which is  why the initial angle should be increased to get large
 kinetic energy.
  The most important thing in the kinetic chain process is  the
  release time. Figure \ref{FRAngVelLength} shows the change
  of the release time for different $l_3$. Considering the longer triple
  pendulum with $l_3 = 0.85 m$, the angular velocity of the second
  rod just before releasing the third rod is greater than the shorter triple
  pendulum with $l_3 =0.65 m$. The releasing time for the third rod of
   the longer triple pendulum is earlier than that of the shorter triple pendulum.
  If we check the time from $t=0$, the acceleration time for the
  shorter triple pendulum is shorter than that of the longer triple
  pendulum.

\begin{figure}[htbp]
\centering
\includegraphics[width=5cm]{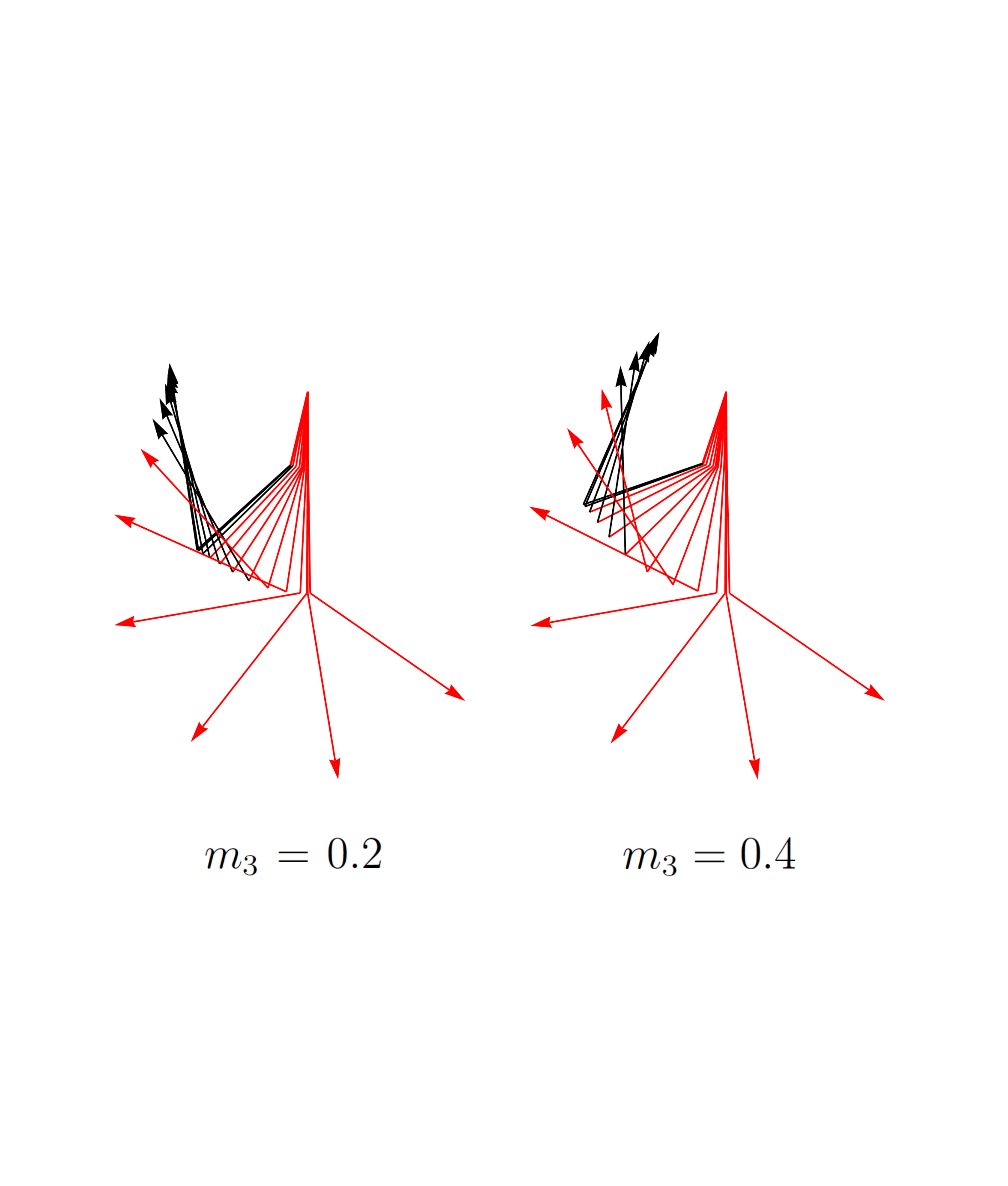}
\caption{The traces of three pendulums for different mass $m_3$ with
keeping other conditions unchanged. The shots are taken at equal
time interval from the starting time. As the length $m_3$ increases
the initial angle of the third rod should be increased. The red rods
are under the force and the black rods  remain fixed without
generating Hooke's force. } \label{FRTraceMass}
\end{figure}

 Figure \ref{FRTraceMass} shows the effect of changing the third
 body. For the lighter triple pendulum with $m_3 =
 0.2 kg $, the initial relative angles among three rods is smaller than
those for the heavier triple pendulum with $m_3 =0.4 kg$. The trend
is almost same considering the shorter and longer triple pendulum.
The reason is related to the kinetic energy of the third rod at
$t=0$. In our simulation, we fixed the angular velocity of the third
rod as  $\dot{\theta_3} = 4 rad/s$, but the kinetic energy of the
third rod is $ \frac{1}{2} m_3  (l_3 \dot{\theta_3} )^2 $, and the
kinetic energy comes from the potential energy which is a function
of the three relative angles among the three rods.

\begin{figure}[htbp]
\centering
\includegraphics[width=5cm]{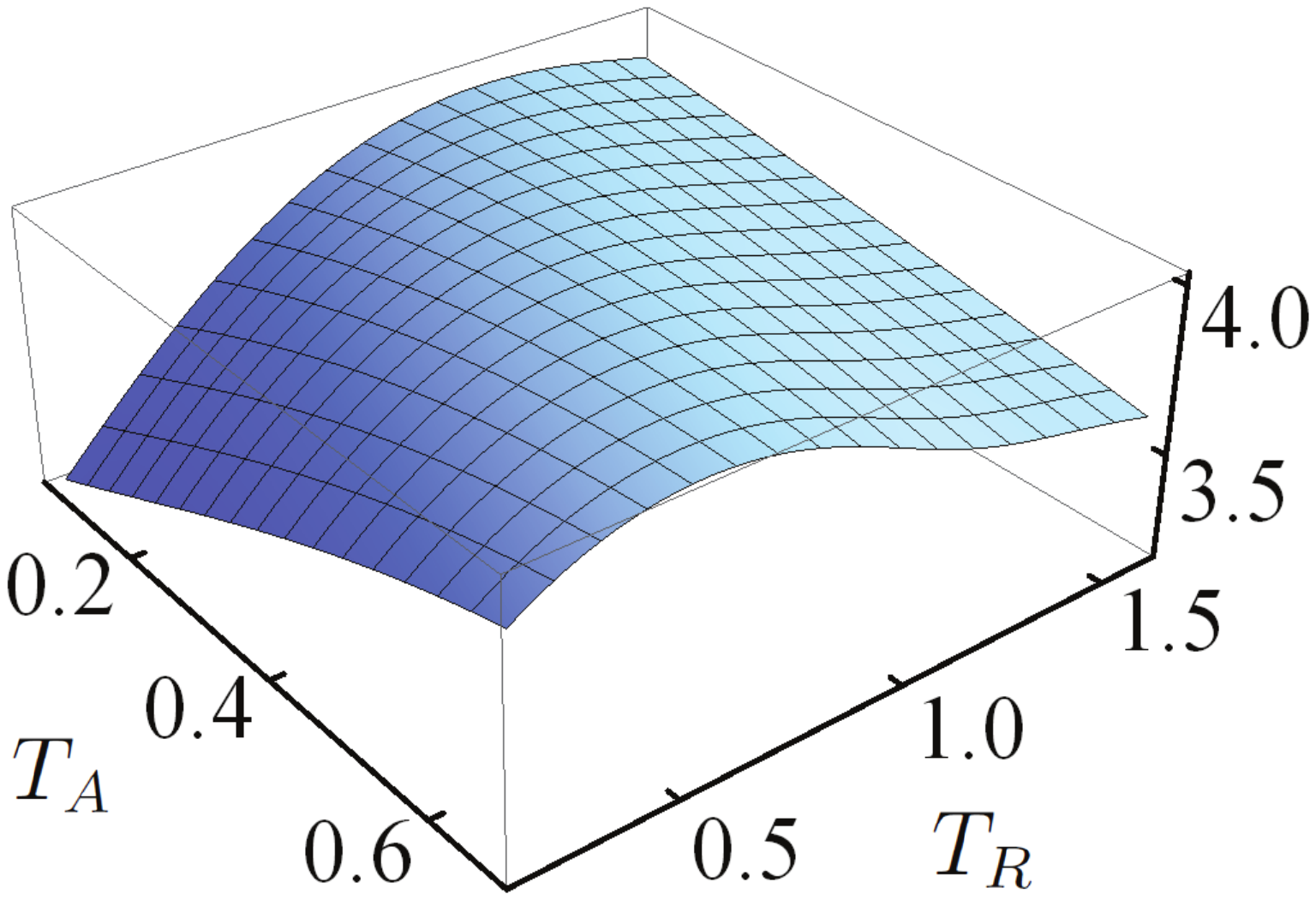}
\caption{ The maximum angular velocity of the third rod as a
function of the release times $T_A$ and $T_R$, where $T_A$ is the
holding time for the second rod, and $T_R$ is the holding time for
the third  rod} \label{FT1T2Change}
\end{figure}

Figure \ref{FT1T2Change} shows the maximum velocity of the third rod
as a function of the release times $T_A$ and $T_R$, where $T_A$ is
the holding time for the second and the third rod, and $T_R$ is the
holding time for the third rod. For the original triple pendulum,
$T_A = -1.89 s - (-2.24)s = 0.35 s$, and $T_R = -1.04 s - (-1.89) s
= 0.85 s$. As shown in  Fig. \ref{FT1T2Change}, the maximum velocity
cannot be increased as $T_A$ and $T_R$ changes from the value $T_A
=0.35 s$ and $T_B = 0.85 s$. That is simply because the condition we
found is the most efficient condition in which the initial potential
energy is totally converted into the kinetic energy of the third
rod. We can only see how fast the maximum velocity changes as the
release times change.

We also find the condition that the initial relative angles among
three rods for a given angular velocity of the third rod as a
function of $m_3$ and $l_3$.
\begin{figure}[htbp]
\centering
\includegraphics[width=5cm]{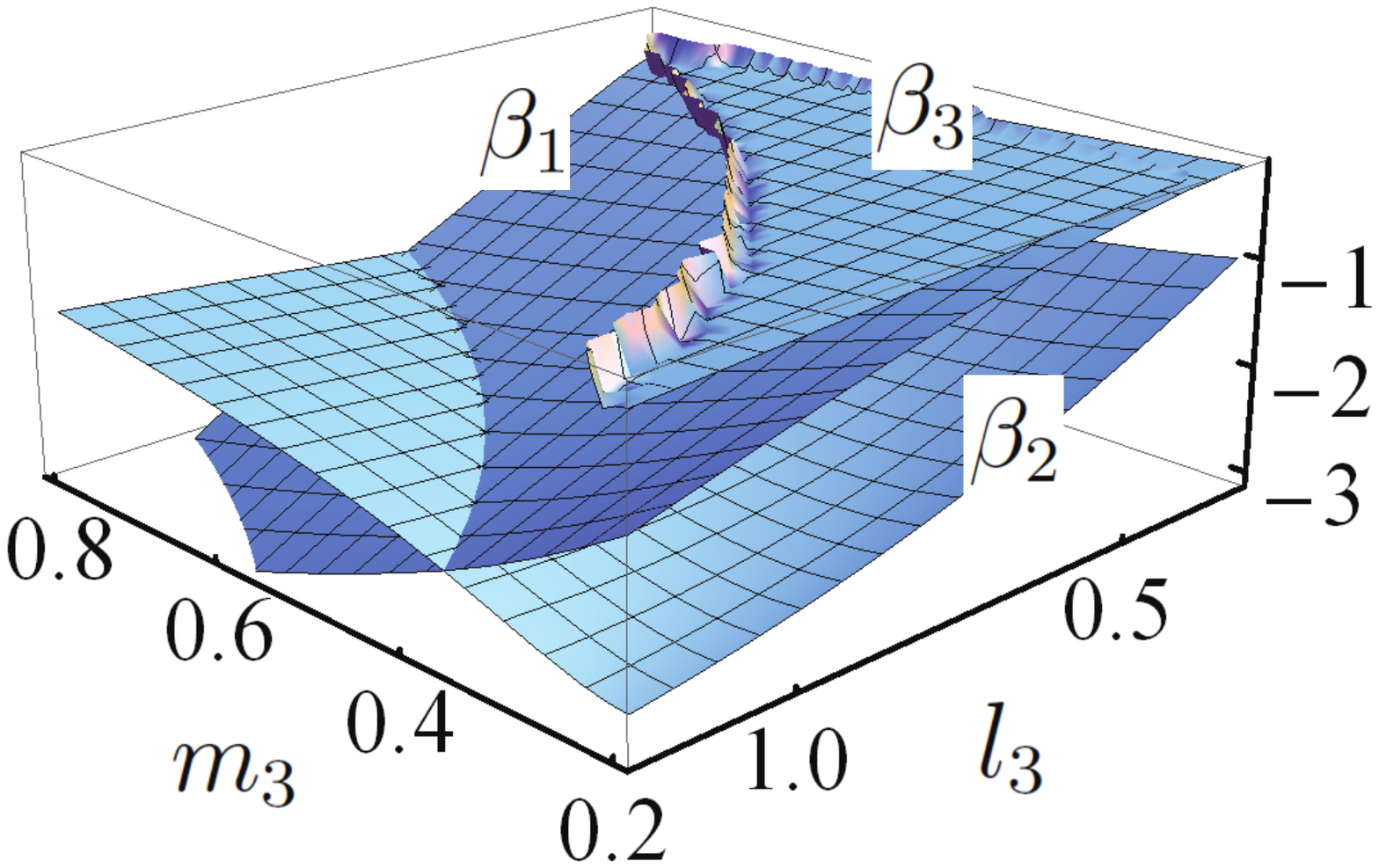}
\caption{Initial angle of $\beta_1$, $\beta_2$, $\beta_3$ as a
function of the mass of the third body($m_3$) and length of the
third rod($l_3$). } \label{FB1B2B3angle}
\end{figure}
The upper-most surface legend by $\beta_3$ is the initial angle of
the first rod. But $\beta_3$ has no negative  solutions for the
entire $m_3$ and $l_3$ space in Fig. \ref{FT1T2Change}. For a small
$l_3=0.5 m $, the initial angle $\beta_3$ has solutions for $0.2 kg
< m_3 <0.7 kg $, but as $l_3$ increases, the initial angle $\beta_3$
has solutions for light mass. Actually there are some solutions for
positive $\beta_3$. In other words, the triple pendulum starts to
moves initially from positive $\beta_3$ and goes back to the
negative angle  with negative angular velocity then returns to the
original direction with positive angular velocity.
 For a given length $l_3$, the initial relative angle $\beta_2$
decreases as $m_3$ decreases, but the initial relative angle
$\beta_1$ increases as $m_3$ decreases. For a given mass $m_3$, the
initial relative angle $\beta_1$ increases as $l_3$ increases, but
the dependence of the initial relative angle $\beta_2$ is complex,
$\beta_2$ increases for light $m_3$ and $\beta_2$ decreases and
finally increases for heavy $m_3$.
 Until now we have studied the kinetic chain process that transfers
 initial potential energy into the kinetic energy of the third body
 using proper time delays. Controlling the time sequence in a triple
 pendulum is an essential element to make a perfect kinetic chain
 process until now.

\section{Triple pendulum without time delay}

\begin{figure}[htbp]
\centering
\includegraphics[width=5cm]{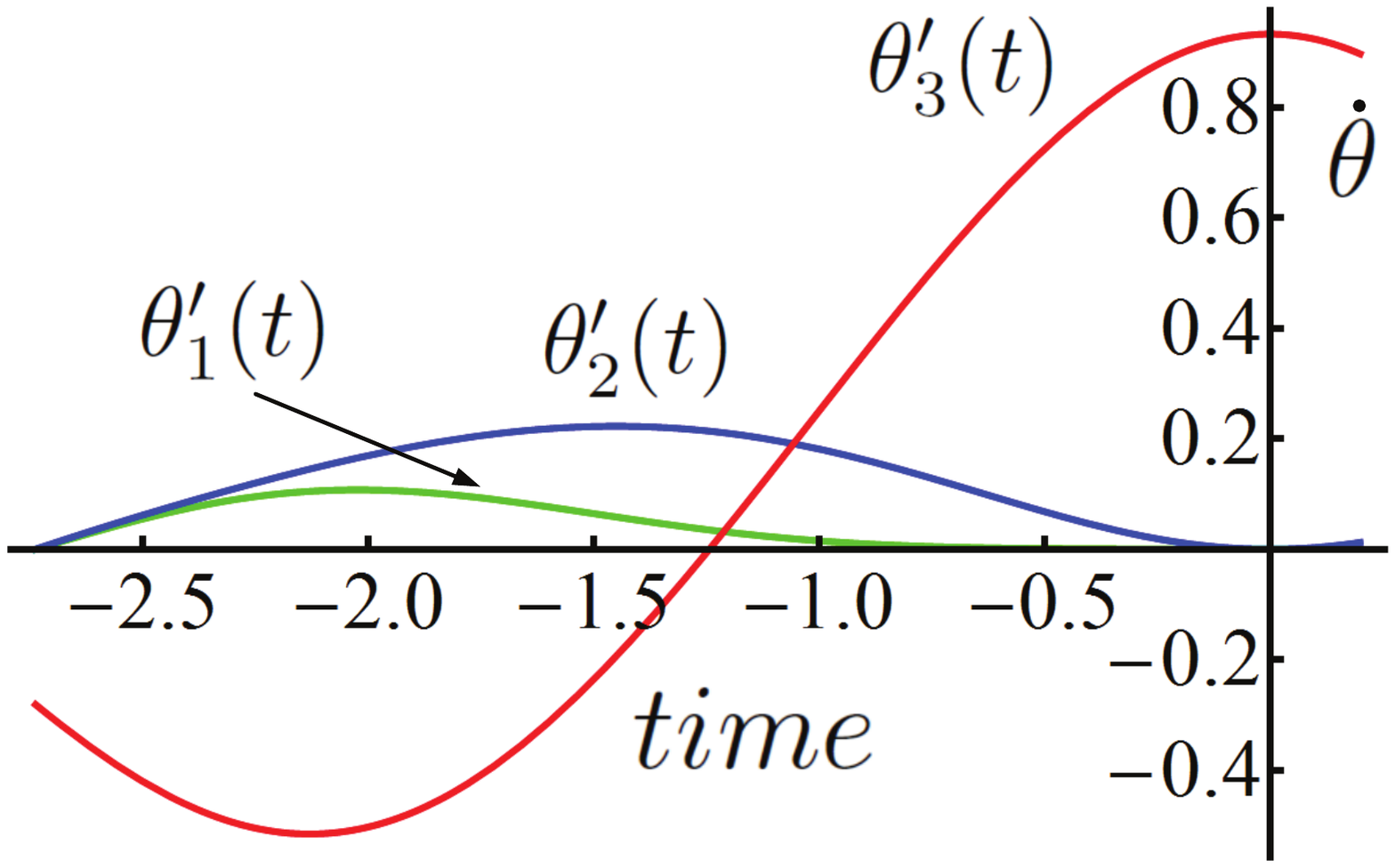}
\caption{Angular velocities of the three rods. The three rods move
from the beginning  without delay, the initial angular velocities of
the first and the second rod are zero and the initial angular
velocity of the third rod is negative. As time passes,   the angular
velocity of the third body changes,  and it has its maximum  $
\dot{\theta_3} = 0.93 rad/s $ at $t=0$. } \label{FTnodelay9}
\end{figure}

Section III, the springs between rods are not simultaneously
released, but in proper time sequence in order to make a perfect
kinetic chain process.  However, if we want to transfer all the
potential energy to the kinetic energy of the third rod, another
solution can be found. Figure \ref{FTnodelay9} shows the solution,
where we plot the angular velocities of the three rods. The three
rods start to move without any time lag from the beginning. At
first, the angular velocities of the first and the second rod are
zero, and the angular velocity of the third rod is negative. The
initial angles of the three angles are $\theta_1 = -6.8 ^{\circ} $,
$\theta_2 = -20.0 ^{\circ} $, $\theta_3 = -8.7 ^{\circ}$. The first
and the second rods move forward, and the angular velocities
increase at first, and then go to zero at $t=0$. On the other hand,
the third rod  starts to move backward, and  turns its direction.
The final velocity of the third rod is $ \dot{\theta_3} = 0.93 rad/s
$.
 If we consider the kinetic chain as a mechanism that efficiently transfers the potential
 energy into kinetic energy, the kinetic chain we found here does not
 require a time lag. The difference from the result in Section
 III is the initial condition. In Section III, the initial
 conditions for the three angular velocities are zero. If we release
 these restrictions, we have a new solution as in Fig.
 \ref{FTnodelay9}, where the initial angular velocity of the third
 rod is not zero, but has a negative value. At $t=0$, the angular
 velocity of the third rod is less than $ 1 rad/s$, which is smaller
 than $ 4 rad/s$. Actually there is no systematic process for arbitrary $\dot{\theta_3}$.
 On the contrary, in Section III, we can systematically find the
 initial conditions that gives perfect energy transfer for a given
 $\dot{\theta_3}$.

\begin{figure}
\includegraphics[width=5cm]{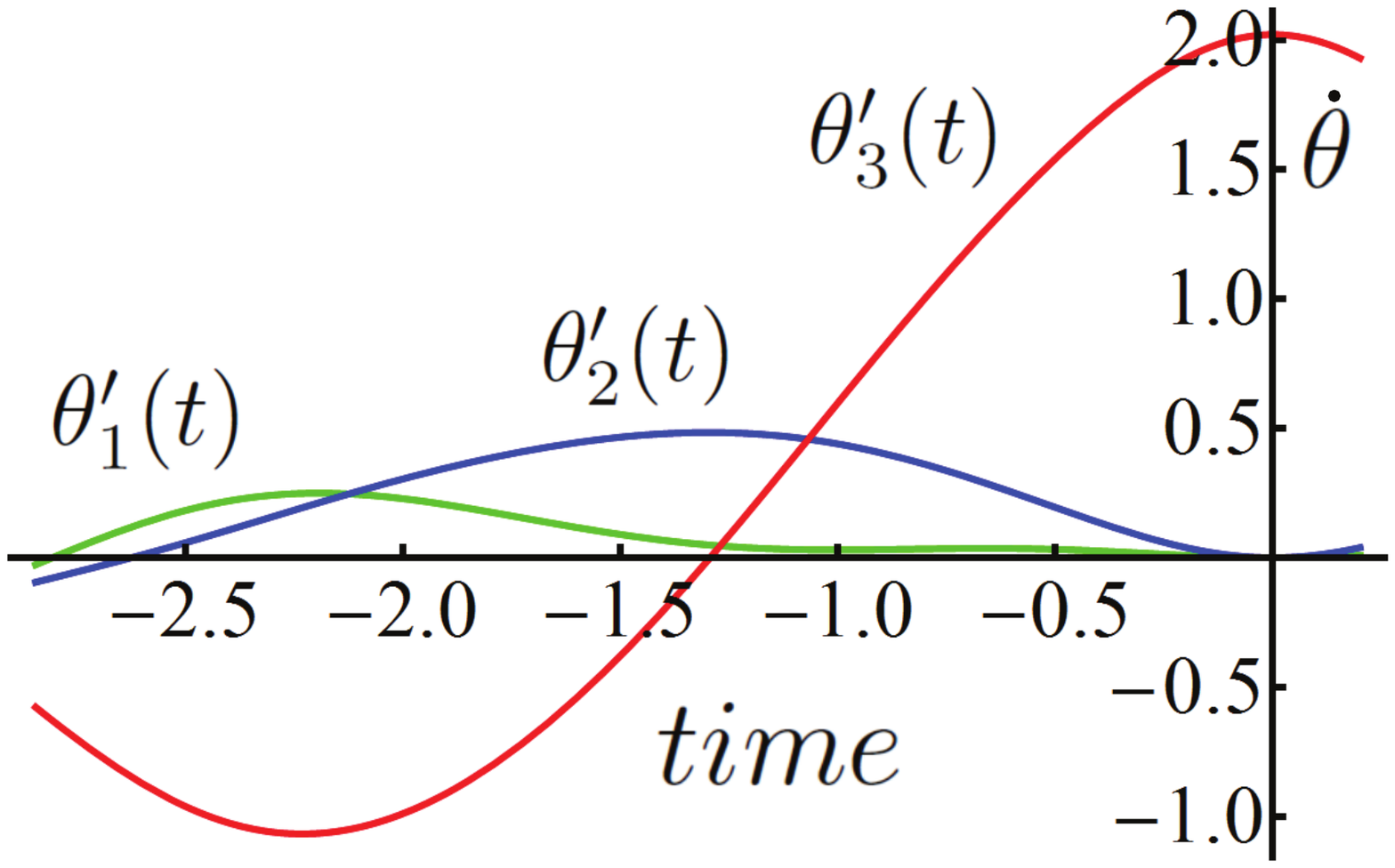}
\caption{Angular velocities of the three rods. The three rods move
from the beginning  without delay, the initial angular velocities of
the first and the second rod are almost zero, and the initial
angular velocity of the third rod is negative. As time passes, the
angular velocity of the third body changes, and it has its maximum
$\dot{\theta_{3}} = 2.0 $ at $t=0$.} \label{Fang2}
\end{figure}

In order to increase the angular velocity of the third rod, we try
to find new solution using a minimizing algorithm.  In Fig.
\ref{Fang2}, the final angular velocity of the third rod is  $
\dot{\theta_3} = 2.0 rad/s $. But the initial angular velocities of
the two rods are not exactly zero, the angular velocities are
$\dot{\theta_1}(0)= - 0.03 rad/s $ and $\dot{\theta_2}(0)= - 0.09
rad/s $, respectively. First, the angular velocity of the first rod
reaches its maximum, and decreases to zero, and at that time, the
angular velocity of the second rod gets its maximum velocity. When
the angular acceleration of the second rod turns to negative, the
angular velocity of the third rod changes from negative value to
positive velocity. At $t=0$, the rod obtains the maximum value $
\dot{\theta_3} = 2.0 rad/s $, and the angular velocities of the
first and the second rod are zero.

\begin{figure}
\includegraphics[width=5cm]{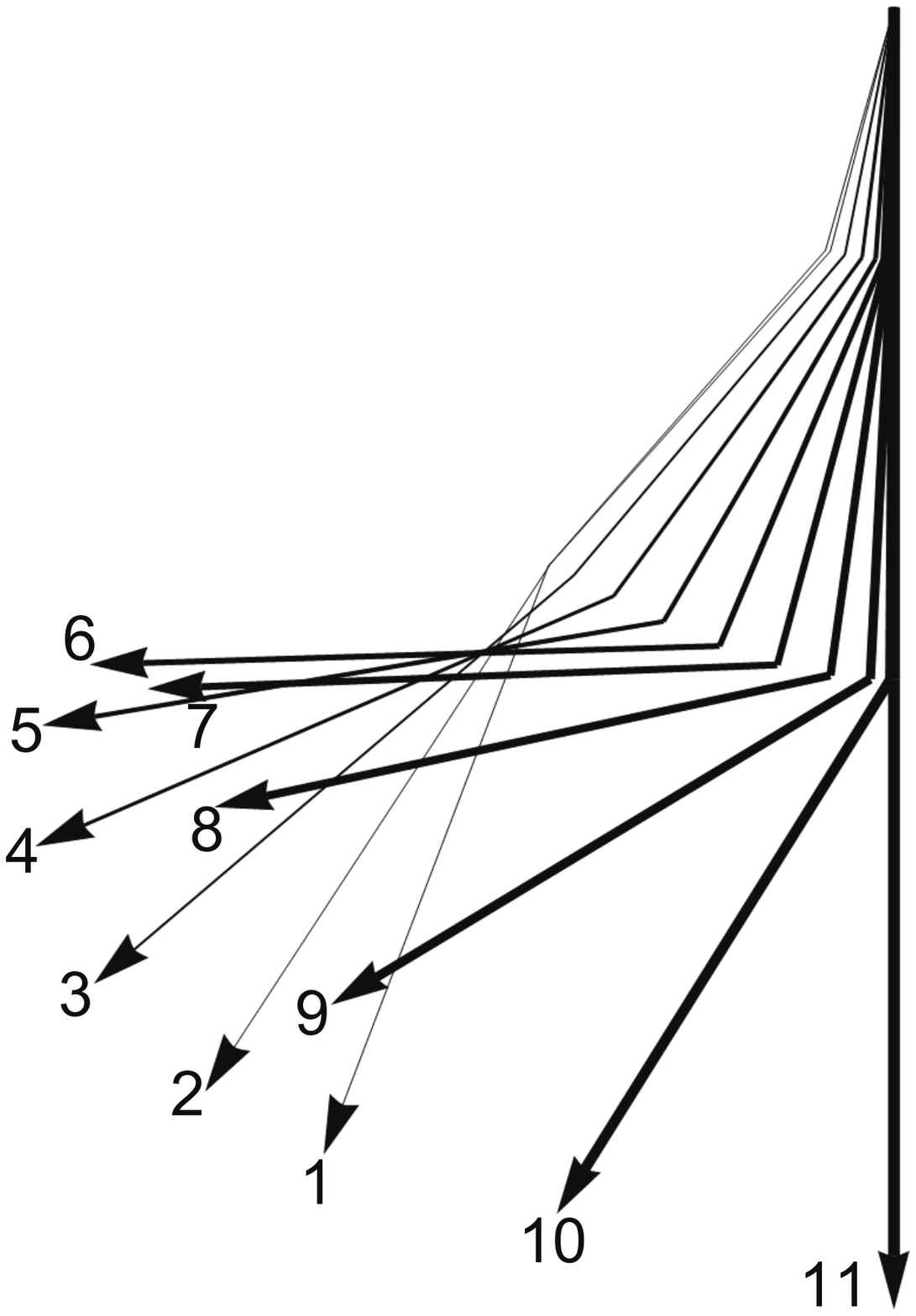}
\caption{The trajectory of the three rods from the beginning time
until the contact time. The third rod moves backward from the first
shot to the sixth shot. After the seventh shot the third rod rapidly
moves forward, and has its maximum angular velocity at the 11th
shot.} \label{FShots2}
\end{figure}

  Figure \ref{FShots2} shows the trajectory of the three rods
from the time beginning until the contact time. At the first shot
(1) in Fig. \ref{FShots2}, the initial angles of three rods are
$\theta_1 = -15.7 ^{\circ}$, $\theta_2 = -41.3 ^{\circ}$ $\theta_1 =
-20.9 ^{\circ}$, respectively.  As shown in this figure, the third
rod moves backward from the first shots to the sixth shot(1-6).
After the seventh shot the third rod rapidly moves forward and has
its maximum angular velocity on the 11th shot at $t=0$. At $t=0$,
the angles of the three rods are all zero.

\begin{figure}
\includegraphics[width=5cm]{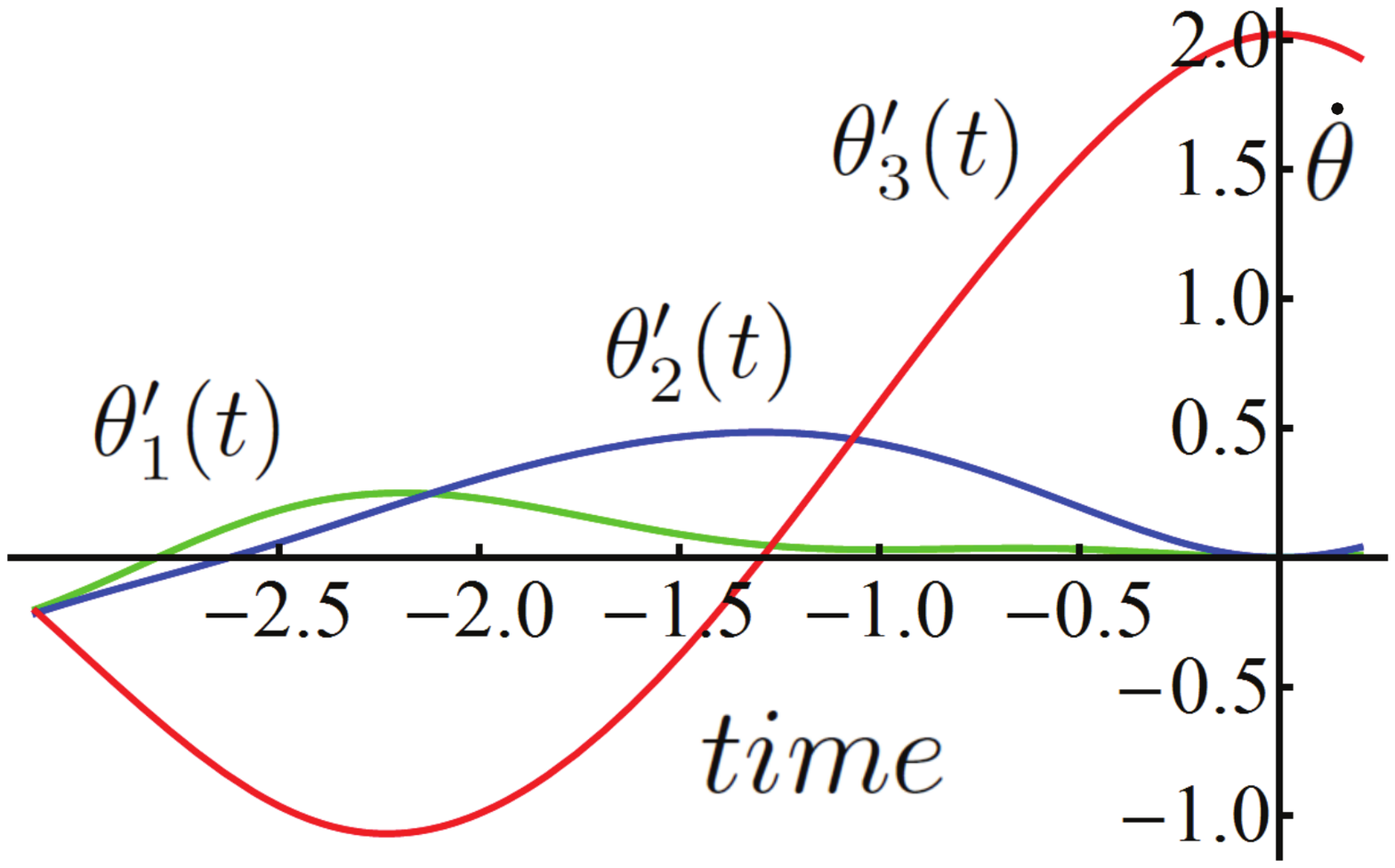}
\caption{Angular velocities of the three rods. The three rods move
from the beginning without delay, the initial angular velocities of
the three rods are the same and negative. After a short time, the
angular velocities of the first and the second rod change into
positive value. The third rod moves backward, changes its direction,
and get its maximum angular velocity  $\dot{\theta_{3}} = 2.0 $ at
$t=0$.} \label{FAngShotsB}
\end{figure}

If we set the initial condition that the three angular velocities
are all the same, there can be another solution that gives efficient
energy transfer.  The initial angular velocities of the three rods
are $\dot{\theta_1}(0)= \dot{\theta_2}(0)= \dot{\theta_3}(0)= -
0.2rad/s $, respectively. In this case, yhr three rods in the triple
pendulum first move together with the same angular velocity.
 As in Fig. \ref{FAngShotsB}, first the angular velocity of the
  first rod changes from the negative
  to the positive, and decreases to be zero,
and  the angular velocity of the second rod also changes from
negative to positive and reaches its maximum velocity. When the
angular acceleration of the second rod turns to negative, the
angular velocity of the third rod changes from the negative value to
the positive velocity. At $t=0$, the angular velocity of the third
rod reaches the maximum value $ \dot{\theta_3} = 2.0 rad/s $, and
the angular velocities of the first and the second rod are zero.

\begin{figure}
\includegraphics[width=5cm]{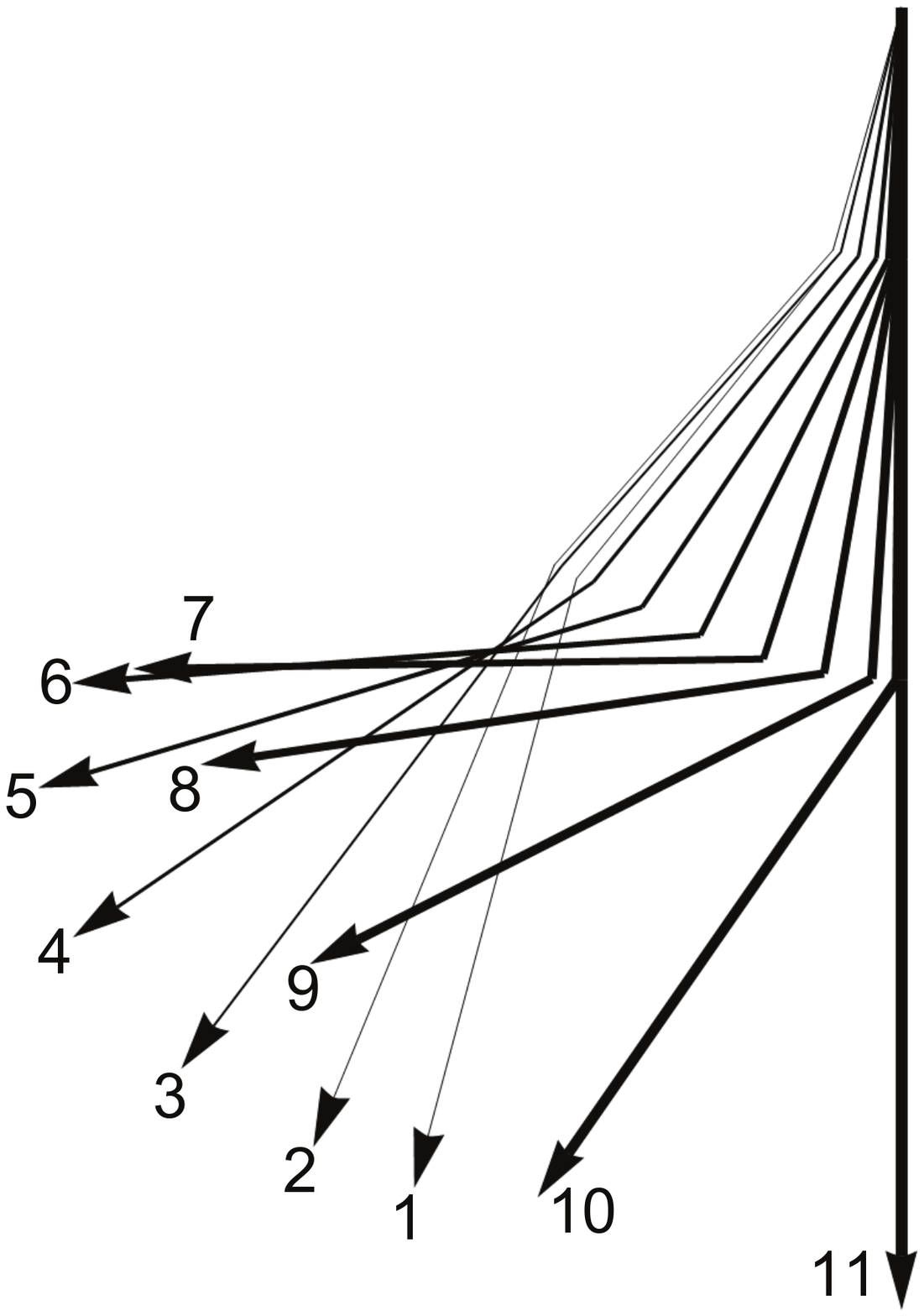}
\caption{The trajectory of the three rods from the time beginning
till the contact time. As shown in this figure, the three rods moves
backward at first. The first and the second rod moves to back ward
till the second shot and they start to move forward from the third
shot. The third rod moves backward till the seventh shot, and it
begin to move forward and has  its maximum angular velocity at the
11th shot.} \label{FShotsB}
\end{figure}

Figure \ref{FShotsB} shows the trajectory of the three rods from the
beginning time until the contact time. At the first shot (1) in Fig.
\ref{FShotsB}, the initial angles of the three rods are $\theta_1 =
-14.0 ^{\circ}$, $\theta_2 = -39.0 ^{\circ}$, $\theta_3 = -14.9
^{\circ}$, respectively.  As shown in this figure, the third rod
moves backward from the first shot to the sixth shot(1-6). After the
seventh shot, the third rods rapidly moves forward, and has its
maximum angular velocity at the 11th shot at $t=0$. At $t=0$, the
angles of the three rods are all zero.

In this section, we introduced a new method to transfer initial
energy including the potential energy of three bodies in a triple
pendulum into the kinetic energy of the third body on the third rod.
This method does not require a time delay which is essential in a
kinetic chain process.

\section{Conclusion and Discussion.}

  The pendulum model is applied to baseball, tennis, and
golf. In particular, the swing pattern using the double pendulum was
studied to maximize the angular velocity of a hitting rod, such as a
racket, bat, or club. The time-lagged torque  effect for the double
pendulum system was studied,  and the speed of the rebound ball can
be increased not by applying time-independent constant torques on
the  first rod and the second rod, but by holding  the racket for a
short time without enforcing a torque and with subsequent
application of torque. This mechanism is related to the well known
kinetic chain process in a swing pattern system.

 In this article, we show the most efficient way to
   transfer input potential energy to the kinetic energy of a racket
   or bat based on the kinetic chain process.
    Introducing the kinetic chain process, we first studied
    a two coupled harmonic  oscillator at first.
   We analytically showed the method to find the
  most efficient way to convert the initial potential energy stored in two
  springs into the kinetic energy of the second body.
  After releasing  the first spring, the two bodies are fix together for the time being.
  Then the two bodies move together with
  a fixed distance between them for a short time $\tau$.
  At $t=\tau$, we release the second spring, then the second body
  absorbs extra kinetic energy from the second spring.  Controlling
  the initial distance of the two springs and $\tau$, the potential energy  initially stored in the
  two springs  can be totally converted into  the kinetic energy of the second body.
  We showed how to find the release time $\tau$ and the
  initially pressed distance of the two springs, in order to get the
  most efficiently converted kinetic energy. This method is the most
  efficient kinetic chain process in a two coupled harmonic
  oscillator.

   We showed how to numerically find the
  most efficient way   to transfer the initial potential energy into the
  kinetic energy of the third mass. We assumed the force on the
  triple pendulum is only the Hooke's force related to the relative
  angles between the rods, as in Fig. \ref{F3Pendulum}. Setting the
  initial angles of the three rods, we can find the  most efficient
  way to transfer the initial potential energy into the kinetic energy of the third body on the third rod
  by adjusting the release times $T_A$ and $T_B$. We are especially interested in the
   length of the third rod and the mass of the third rod, which are related to the racket
  system. The racket system may change if we use different rackets or
  bats or golf clubs. Although our numerical data are not well
  matched to an actual system, the trend can be applied.

 In the kinetic chain process, we usually point out the sequential
 application of torque or force for each rod in a multiple pendulum
 model. However,   there is a  new
 method that gives perfect energy transfer from the initial
 conditions into the kinetic energy of the third body on the third
 rod in a triple pendulum system. The new method did not use the
 sequential time delays that are essential in a kinetic chain
 process. However, if we plot the time dependent angular
 velocities of three bodies as in Figs.
 \ref{FTnodelay9}, \ref{Fang2}, \ref{FAngShotsB}, the trend of the angular
 velocities of the first and the second bodies are the  same as the
 result in Fig. \ref{FRAngVelLength}. As the angular velocity of the
 first rod decreases, the velocity of the second rod increases. However, the
 trend of the angular velocity of the third rod is totally different. In the
 new  method in Section IV, the angular velocity of the third body
 is negative at first, then after the acceleration of the second rod becomes negative,
 the angular velocity of the third body changes sign.  Finally the velocity of the third rod reaches
 its maximum when the angular velocities of the other rods are zero.

  Of course, we used a simple pendulum model with only Hooke's
  force, so the results are not directly applicable to the actual swing
  process using a racket, bat, or golf club. However, we
  systematically showed how to find the most efficient way to
  transfer initial potential energy to the kinetic energy of the
  third rod in a kinetic chain process. Furthermore, we showed an
  new method to transfer the initial energy to
  the final kinetic energy without a time lagged process, which is required
  in a kinetic chain process.

   In av actual stroke process, the motion occurs in three dimensions and
   the  magnitudes of the forces and torques depend on the
muscle shape and movement. We did not include any bio-mechanical
information such as pronation.  The numerical data may also not be
suitable for some players. However, our study on the kinetic chain
process using a triple pendulum system for a stroke provides some
insights into attaining an efficient way to transfer body energy to
a stroked ball.

\end{document}